\documentclass[aps,prx,twocolumn,english,superscriptaddress,longbibliography]{revtex4-2}

\usepackage{boldline}
\usepackage{bbold}
\usepackage{xcolor}
\usepackage{graphicx,amsmath}
\usepackage{amssymb}
\usepackage{mathrsfs}
\usepackage{booktabs}
\usepackage{array}
\usepackage{multirow}

\newcommand{\sugg}[1]{\textcolor{red}{#1}}
\newcommand{\corr}[1]{\textcolor{blue}{#1}}
\newcommand{\pad}[1]{\textcolor{green}{#1}}

\newcommand{\PL}[1]{\textcolor{purple}{#1}}

\let\corr\relax
\let\PL\relax
\let\sugg\relax
\let\pad\relax

\begin{document}

\title{Quantum-limited detection of arrival time and carrier frequency of time-dependent signals}

\author{Patrick Folge}\thanks{These authors contributed equally}
\affiliation{Paderborn University, Integrated Quantum Optics, Institute for Photonic Quantum Systems (PhoQS), Warburger Stra{\ss}e 100, 33098 Paderborn, Germany}

\author{Laura Serino}\thanks{These authors contributed equally}
\affiliation{Paderborn University, Integrated Quantum Optics, Institute for Photonic Quantum Systems (PhoQS), Warburger Stra{\ss}e 100, 33098 Paderborn, Germany}

\author{Ladislav Mi\v{s}ta, Jr.}
\affiliation{Department of Optics, Palack\' y University, 17.
listopadu 12,  779~00 Olomouc, Czech Republic}

\author{Benjamin Brecht}
\affiliation{Paderborn University, Integrated Quantum Optics, Institute for Photonic Quantum Systems (PhoQS), Warburger Stra{\ss}e 100, 33098 Paderborn, Germany}

\author{Christine Silberhorn}
\affiliation{Paderborn University, Integrated Quantum Optics, Institute for Photonic Quantum Systems (PhoQS), Warburger Stra{\ss}e 100, 33098 Paderborn, Germany}

\author{Jaroslav \v{R}eh\'{a}\v{c}ek}
\affiliation{Department of Optics, Palack\' y University, 17.
listopadu 12,  779~00 Olomouc, Czech Republic}

\author{Zden\v{e}k Hradil}
\affiliation{Department of Optics, Palack\' y University, 17.
listopadu 12,  779~00 Olomouc, Czech Republic}


\begin{abstract}
\sugg{ Precise measurements of both the arrival time and carrier frequency of light pulses are essential for time–frequency–encoded quantum technologies. Quantum mechanics, however, imposes fundamental limits on the simultaneous determination of these quantities.
In this work, we derive and experimentally verify the quantum uncertainty bounds governing joint time–frequency measurements. We show that when detection is restricted to finite time windows, the problem is naturally described by a quantum rotor, rendering the commonly used Heisenberg uncertainty relation inapplicable. We further propose an optimal detection scheme that saturates these fundamental limits.
By sampling the Q-function, we demonstrate the reconstruction of the Wigner function beyond the harmonic oscillator. Using an experimental implementation based on a quantum pulse gate, we confirm that the proposed scheme approaches the ultimate quantum limit for simultaneous time–frequency measurements.
These results provide a new framework for joint time–frequency detection with direct implications for precision measurements and quantum information processing.
}
\end{abstract}
\maketitle


\date{\today}


\section{Introduction}

Time occupies a unique and pivotal role in physics, bridging diverse disciplines and prompting numerous conceptual questions that demand answers \cite{Altaie_22}. In quantum mechanics, time governs the evolution of quantum systems, acting as a fundamental parameter that defines their dynamical transformations. At the same time, time also plays a central role as a quantity to be estimated from measurement outcomes, where it enters as a parameter characterizing the quantum state. This viewpoint lies at the heart of estimation and detection theory, which, although rooted in classical statistical signal analysis, was systematically formulated in the quantum domain nearly fifty years ago in the seminal work of C.~W.~Helstrom \cite{Helstrom_76}. His framework continues to underpin modern developments in quantum metrology and quantum technologies.

A particularly challenging and long-debated problem in this context is the efficient and simultaneous estimation of the arrival time and the carrier frequency of a time-dependent quantum signal at the ultimate precision limits. This difficulty originates from the distinctive structure of the time--frequency (TF) phase space, which is fundamentally different from the canonical phase space of position and momentum defined over unbounded domains. In realistic measurements, causality and finite detection times inevitably restrict the observation window to a finite temporal interval, which profoundly modifies the attainable precision limits.

The precision with which two quantities can be simultaneously estimated is intimately connected to their commutation relations. The archetypal example is Heisenberg’s uncertainty principle for the position $x$ and momentum $p$ of a quantum particle \cite{Heisenberg_27}, which is formalized by the Robertson inequality $\Delta x\,\Delta p \ge \hbar/2$ \cite{Robertson_29}. This bound is saturable by Gaussian states and follows from the Heisenberg algebra of non-commuting operators, $[x,p]=i\hbar\openone$, reflecting the fact that position and momentum are related by a continuous Fourier transform. The same mathematical structure underlies many cornerstone concepts of quantum optics and quantum information science, including coherent and squeezed states \cite{Glauber_63,Yuen_76}, Einstein--Podolsky--Rosen correlations \cite{Einstein_35}, the Arthurs--Kelly theory of joint measurements \cite{Arthurs_65}, and phase-space representations of quantum states introduced by Wigner, Husimi, and Glauber \cite{Wigner_32,Husimi_40,Glauber_63}.

Difficulties arise, however, when attempting to apply this framework to the estimation of the arrival time and carrier frequency of a time-dependent quantum signal, such as a pulsed single photon. These two parameters naturally characterize so-called time--frequency plaquettes used for encoding quantum information \cite{Helstrom_76}. While time itself is not a quantum observable in nonrelativistic quantum mechanics, its optimal estimator can be consistently represented by a quantum operator within estimation theory. Crucially, in the presence of a finite detection time window typical of realistic experiments, the Heisenberg algebra is no longer an appropriate description of the time--frequency relationship. Although standard Heisenberg-type uncertainty relations and chronocyclic phase-space descriptions yield accurate results when the detection window is much larger than the pulse duration \cite{Brecht2015,Bhattacharjee2025}, they do not provide tight or saturable bounds for both complementary variables under finite-time constraints.

\sugg{ The central role in our formulation is therefore played by \emph{saturable} TF uncertainty relations adapted to finite temporal domains. Since the detection time window is explicitly restricted to a finite interval, the commonly quoted Heisenberg-type relations $\Delta t\,\Delta\omega \ge 1/2$, which are saturated by Gaussian states defined over an infinite time axis, is no longer tight in this regime. Alternative approaches include Mandelstam--Tamm-type bounds \cite{Mandelstam1945}, which provide useful limits on dynamical evolution but are not generally saturable for finite-support signals. Our approach is conceptually closer to methods developed in harmonic analysis for sequences with minimal TF uncertainty \cite{Parhizkar2015}, reformulated here within a quantum-mechanical framework. A particularly important role is played by Cramér--Rao inequalities involving time and frequency parameters. For clarity, the reader is referred to Table~I, which summarizes and contrasts different classes of uncertainty relations and highlights that only rotor-like uncertainty relations admit saturable bounds and can be consistently extended to the simultaneous estimation of non-commuting variables.}

\sugg{ In this article, we develop and experimentally demonstrate a method for optimal simultaneous detection of the arrival time $t_0$ and carrier frequency $\omega_0$ of an optical pulse over a finite time interval. These two parameters are singled out because they admit a well-defined and saturable quantum precision bound in the presence of a finite detection window. Although the measurement scheme is tomographically complete and, in principle, allows reconstruction of the full quantum state, our focus is explicitly metrological: we aim to identify optimal measurement strategies and ultimate precision limits for estimating the physically relevant displacement parameters $t_0$ and $\omega_0$ under realistic constraints. }

\sugg{ Our results are obtained in several steps. First, we derive saturable uncertainty relations under finite-time constraints using a formulation based on an effective expansion of the field that naturally leads to the Euclidean symmetry group $E(2)$ associated with a quantum rotor. The resulting uncertainty relations are saturated by von~Mises states, which play a role analogous to squeezed states in the Heisenberg algebra. In the second step, the formalism is extended to simultaneous TF estimation, enabling optimal quantum tomography of the pulse. Experimentally, the simultaneous measurement is implemented using a quantum pulse gate, where a pulse prepared in a von~Mises state is projected onto a set of displaced von~Mises modes in the TF domain, effectively sampling the corresponding $Q$ function. The measured data demonstrate that both state preparation and detection operate close to the ultimate quantum limits imposed by TF uncertainty. }

The paper is structured as follows. In Sec.~\ref{sec_Theory}, we apply the quantum rotor formalism to signals defined on a finite time interval and derive the corresponding saturable uncertainty relations. In Sec.~\ref{sec_Estimation}, we address the estimation of the carrier frequency in a finite temporal domain. Section~\ref{sec:simultaneous_measurement} extends the theory to optimal simultaneous TF detection. The experimental implementation is described in Sec.~\ref{sec_Experiment}, with results presented in Sec.~\ref{sec_Results}. Finally, Sec.~\ref{sec_Conclusion} summarizes our conclusions.

\section {TF variables in the formalism of quantum rotor }\label{sec_Theory}
\begin{figure*}
    \centering
    \includegraphics[]{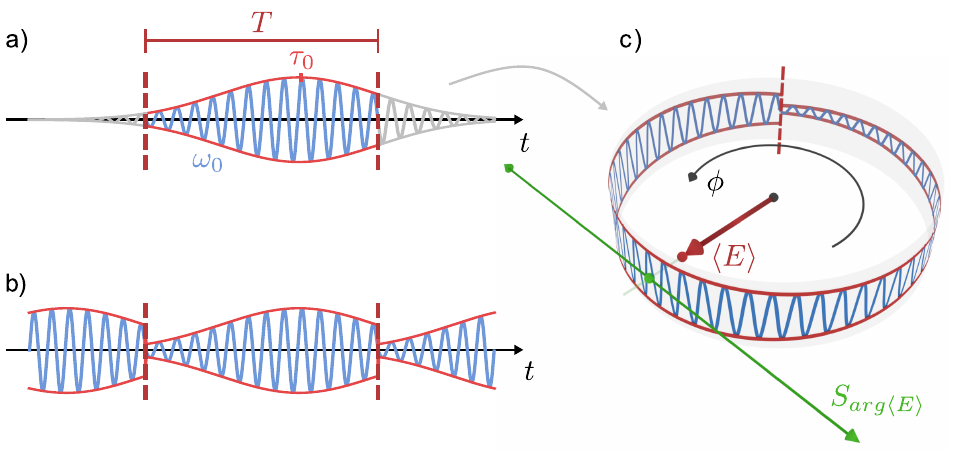}	   
    \caption{(a) We consider a time-dependent quantum signal with an arrival time $\tau_0$ and a carrier frequency $\omega_0$ on a finite time interval $T$. (b) This signal can be periodically extended outside the time-interval, which motivates the mapping to a unit circle. (c) By mapping to the unit circle, we introduce an angular variable $\phi$ that describes the signal. The spread of the signal is characterized by the standard deviation $\Delta S$ of the rotated sine operator (\ref{S}). For more information, see the text.}
    \label{fig:pulse}
\end{figure*}

\PL{We consider a time-dependent signal described by its complex amplitude $\psi(t)$ in} the finite time interval $T$ as sketched in Fig. \ref{fig:pulse}. After a periodic extension, the time variable $t$ can be mapped \PL{onto the} dimensionless angular variable $\phi=2\pi\frac{t}{T}$. This allows us to write the signal as a Fourier series

\begin{equation}
    \psi(\phi) = \sum_{l\in\mathbb{Z}}a_l e^{il\phi}
    \label{pulse}
\end{equation}
with coefficients $a_l$.  To maintain consistency with the rotor model, Greek letters will be used to denote the time variable, and renormalization will be applied only when necessary.


\PL{This rotor-like geometry is described by the} Euclidean $E(2)$ algebra\PL{, which} canonically describes the angular momentum \PL{$L = -i\partial_{\phi}$} with integer eigenvalues $l\in\mathbb{Z}$ and eigenbasis $\{|l\rangle\}_{l\in\mathbb{Z}}$, and the unitary shift operator \PL{$E=e^{-i\phi}$}, $E|l\rangle=|l-1\rangle$, with eigenvalues $e^{-i\phi}$ and eigenvectors $\{|\phi\rangle=\sum_{l\in\mathbb{Z}}e^{-il\phi}|l\rangle/\sqrt{2\pi}\}_{\phi\in[-\pi,\pi)}$.

Angular momentum and shift operator satisfy the commutation relation $[E, L] = E$, which is characteristic for a quantum rotor - a periodic quantum system with infinite-dimensional Hilbert state space $L^{2}(-\pi,\pi)$ \cite{Albert_17}. 
 Alternatively and equivalently, the operator $L$ can be interpreted as a self-adjoint extension of the momentum operator $p$ on a finite interval. This nontrivial result in mathematical physics is elegantly and pedagogically explained in \cite{Bonneau_01}.
   The detailed theory of minimum uncertainty states for a quantum rotor was developed in \cite{Mista_22}. A follow-up paper \cite{Mista_24} introduces additional measures for quantifying the saturable uncertainties of general complementary quantum rotor variables. 
   
   \PL{In the following, we will apply the same formalism to time (mapped to $\phi$) and frequency $\omega$ \sugg{(mapped to $l$)}, which will be described by $E$ and $L$, respectively.} Below, we briefly review the key theoretical concepts necessary for understanding our experimental demonstration.

We formulate the theory in the four steps: i) we set the fundamental uncertainty relations that arise from this algebra;  ii) we discuss the relation between discrete and continuous frequency variables; iii) we formulate  the simultaneous measurement of the TF variables; iv) finally, we give a recipe for retrieving the resulting uncertainties from measurement data.



We review here briefly the form   which can be derived from the commutation relation $[E, L] = E$ of the $E(2)$ algebra \cite{Mista_22}

\begin{equation}
    \label{eq:tf_uncertainty}
    \Delta L\,\Delta S\geq \frac12 |\langle E \rangle|,
\end{equation}
where $\Delta X\equiv\sqrt{ \langle(\Delta X)^2\rangle}$ is the standard deviation and
\begin{equation}\label{S}
    S = \frac{1}{2i} \left( e^{i\mathrm{arg}\langle E \rangle}E^\dagger - e^{-i\mathrm{arg}\langle E \rangle} E \right)
\end{equation}
is the sine operator rotated by the state-dependent angle $\mathrm{arg}\langle E \rangle$.
The term  $\Delta S^2$ can be interpreted as a moment of inertia of the mass distributed on the unit circle with respect to an axis in the plane of the circle that connects the center of mass with the origin \cite{Mista_24}, c.f. Fig.~\ref{fig:pulse}(c).  
 $\Delta  S $ quantifies the precision to determine  the time of arrival of the signal linked with the condition $\langle S \rangle = 0. $ \sugg{It means that average time of the signal field $\tau_0$ is defined by the condition $\langle \sin[\frac{2\pi}{T} (t-\tau_0) ] \rangle = 0 $ whereas $\omega_0 = \frac{2\pi}{T}\langle L \rangle.$}
By defining a normalized uncertainty $\sigma = \Delta S/ |\langle E \rangle|$,
this can be rewritten to the final form
\begin{equation}
    \label{eq:normalized_tf_uncertainty}
    \Delta L\,\sigma\geq \frac12.
\end{equation}

Importantly, the uncertainty relations (\ref{eq:tf_uncertainty}) and (\ref{eq:normalized_tf_uncertainty}) are saturated by the states \cite{Bluhm_95,Kastrup_06,Hradil_10}
\begin{eqnarray}
|n,\alpha\rangle=\frac{1}{\sqrt{I_0( 2 \kappa) }}\sum_{l\in \mathbb{Z}} e^{i(n-l)\alpha}I_{n-l} (\kappa)|l\rangle
\label{eq:von_mise_state}
\end{eqnarray}
possessing the von Mises distribution for the angular (time) variable 
$\phi$: $\langle\phi|n,\alpha\rangle  \propto\exp{[ i n\phi +  \kappa \cos(\phi-\alpha)]}.$ 
\PL{For this reason,} these states will be  called von Mises states.
Here $n\in\mathbb{Z}$ is the angular momentum mean, $\alpha\in[-\pi,\pi)$ is an angle, and $I_{n}(z)$ is the modified Bessel function of order $n$ \cite{Watson_44}. The parameter $\kappa\geq0$ represents the spread of the angular variable, which is analogous to quadrature squeezing. A convenient way of \PL{preparing} the shifted von Mises states is as $|n,\alpha\rangle =D(n,\alpha)|0,0\rangle$, where $D(n,\alpha) = e^{-iL\alpha} E^{-n}$ is the displacement operator, which displaces the fiducial state 
$|0,0\rangle\equiv|n=0, \alpha=0\rangle$ in time and frequency. An interesting property of the von Mises states is that their scalar product retains its shape after the discrete-continuous Fourier transformation associated with the $E(2)$ algebra (see Appendix  for details). In this sense, they are the equivalent to Hermite-Gaussian functions in the continuous domain, but on a finite interval. As will be seen further, these states will play a key role in optimal measurements, since their projections will yield information about \PL{the} discrete frequencies $n$ and \PL{the} time variable $\alpha$ here normalized to $2\pi$ window.

\section{Relation between discrete  and  continuous frequencies}
\label{sec_Estimation}

\sugg{ In this section, we analyze TF variables from the perspective of parameter estimation. This requires clarifying the relationship between optical descriptions based on continuous frequency variables and the effective rotor formulation, in which a discrete index naturally appears.}

\sugg{ In the optical rotor formalism, this discrete index can be interpreted as a mode number, analogous to resonator modes, and arises from a Fourier-series representation associated with finite or periodic boundary conditions in time. The resulting discreteness is therefore a property of the effective description of the problem, rather than a statement about the microscopic frequency content of the optical field.} 

\sugg{ Importantly, adopting a quantum-rotor formalism does not imply that the optical signal itself must be generated from discrete frequency components. In the experiment presented below, the optical pulses are synthesized using a continuous frequency spectrum. The rotor-based description is introduced because it provides a natural and consistent framework for treating TF variables and their simultaneous estimation within a finite temporal domain.} 

In mathematical terms,  formulation  involves extension of  the signal  to an infinite interval by setting it to zero outside the given time window.  To maintain a clear physical interpretation, we use a real parameterization of the time window $T$. \PL{We can relate this continuous frequency spectrum $\omega$ to the discrete frequencies $l$ derived from the quantum rotor formalism by describing the state $\psi(t)$ in terms of $\omega$:} 
\begin{eqnarray}
\Psi(\omega)&=&   \frac{1}{\sqrt{2 \pi} }  e^{i\omega \tau_0 }   \int_{0}^{T} {d t}\psi(t)e^{-i\omega t },
\label{FTpulse}
\end{eqnarray}
where    $\tau_0$ is the time of arrival  (mean value of the  time variable). 
This allows one to define the $\omega$- and $t$-basis  as 
\begin{eqnarray}
|\omega\rangle=\frac{1}{\sqrt{2\pi}}  \int_{0}^{T }d t|  t\rangle e^{i\omega t} \\
= \sqrt{\frac{2}{\pi T}} e^{i\frac{\omega T}{2}} \sin{\left(\frac{\omega T}{2}\right)} \sum_{l\in\mathbb{Z}} \frac{|l \rangle }{ \omega -\frac{2\pi  }{T} l } , \\
|t\rangle =   \frac{1}{\sqrt{T}}  \sum_{l\in\mathbb{Z}}   e^{-i\frac{2  \pi}{T} tl}|l\rangle.
\end{eqnarray}
Notice that $\omega$-basis is non-orthogonal but over-complete since 
\begin{eqnarray}
\langle\omega'|\omega''\rangle=   \frac{T}{2\pi }e^{i \frac{T}{2} (\omega'-\omega'') }  \frac{1}{{2\pi}}  {\rm sinc}\left[\frac{T (\omega'-\omega'')}{2}\right],  \\
\int_{-\infty}^{+\infty} d \omega|\omega\rangle \langle \omega| =  \int_0^T dt|t\rangle \langle t| =  \openone_T,
\end{eqnarray}
where  the sinc function is defined as
$ {\rm sinc} x =\frac{\sin x }{x} . $

However, a key challenge arises in \PL{the interpretation of this description}: the probability distribution
$p(\omega|l) =|\langle \omega|l\rangle|^2\propto{\rm sinc}^2[(\omega-\frac{2\pi}{T}l)\frac{T}{2}]$  exhibits a diverging variance, $\Delta \omega^2,$ analogously to the \PL{position after diffraction on a slit} \cite{Rehacek_04}. This highlights the difference between variance and \corr{ estimation precision (linked to Fisher information \cite{Rehacek_04}): }for a sinc-shaped probability distribution, the variance $\Delta \omega^2$ increases with the number of detection events used to evaluate it, while the "center of mass" of the distribution is determined with a precision $\Delta \omega_{est}^2$  
 \begin{eqnarray}
\Delta \omega_{est}^2 \ge  \frac{1}{N F_{\omega}},
\end{eqnarray}
{$N$- number of samples  will be set to $N=1$ in the following, and  } 
\begin{align}
\label{Fisher_omega}
F_{\omega} & =\int d \omega  \frac{[\Psi(\omega)'^* \Psi(\omega) + \Psi(\omega)^* \Psi(\omega)']^2}{\Psi(\omega)^* \Psi(\omega)}\nonumber\\
= & \,\,4\int d\omega |\Psi'(\omega)|^2 +
\int d \omega  \frac{[\Psi(\omega)'^* \Psi(\omega) - \Psi(\omega)^* \Psi(\omega)']^2}{\Psi(\omega)^* \Psi(\omega)},\nonumber\\ 
\end{align}
The abbreviation  $' = \partial_{\omega}$ is used. Since the first term is four times the variance of time, $\Delta t^2,$  and  the second term is non-positive, we recover the \PL{CR inequality
\begin{eqnarray} 
\label{eq:cr}
\Delta \omega_{\text{est}}^2 \Delta t^2 \ge \frac{1}{4}. 
\end{eqnarray}
}
Indeed, this is just an illustration of the optimal estimation (Quantum Fisher Information) of the parameter $ \epsilon $ generated by   $|\psi(\epsilon) \rangle = e^{i \epsilon G} |\psi \rangle$, with $\Delta \epsilon_{QFI}^2 = 4 \Delta G^2$. \PL{Though Eq.~\eqref{eq:cr}} may seemingly resemble an "uncertainty relation", it is not truly one, as the frequency $\omega$ is not measured but estimated, and the time variance is not a proper measure of uncertainty over a finite time interval  serving here rather as a constraint on the Fisher information.

This illustrates the subtle distinction between "detection" and "estimation" strategies. Projections onto frequency eigenstates can be used to construct efficient estimators, such as the maximum likelihood estimator, enabling frequency estimation at the ultimate precision allowed by Fisher information. However, this is not the TF uncertainty relation, even though Fisher information is constrained by the time variance. The time variable cannot be determined using $|\omega \rangle  $ projections, as these projections are mutually unbiased \PL{and satisfy} 
$ | \langle t | \omega \rangle |^2   =  \frac{1}{2\pi}, $ despite the fact that time and frequency are not strictly conjugate variables. The uncertainty relations associated with all the TF variables considered so far are summarized in Table 1.
\begin{table*}
\caption{\label{table}  \pad {Uncertainty relations for  time and frequency. }}
\begin{ruledtabular}
\begin{tabular}{>{\centering\arraybackslash}m{4cm}  m{12cm}} 
   Uncertainty relations & Comments \\ 
\midrule
\multirow{-2.3}{*}{$\Delta t\,\Delta \omega \geq \frac{1}{2} $} & 
Uncertainty for the Heisenberg-like pair $[t, \omega] = i $, which is valid in infinite domains, $t,\omega\in(-\infty,+\infty)$ and saturated by Gaussian states. 
Uncertainties cannot be saturated in the finite time domain, where $\Delta t$ is an inappropriate measure. \\ 
\midrule
\multirow{-3.3}{*}{$ \sigma\,\Delta L\geq \frac{1}{2}$} & 

\PL{Normalized uncertainty for the rotor-like pair  $[E, L] = E$ on a finite time-interval. $\Delta L$ measures the spread of the discrete frequency variable $l\in\mathbb{Z}$ and $\sigma$ measures the spread of the time-variable $\phi\in[-\pi,\pi)$.} The uncertainty is saturated by von Mises states and can be converted to simultaneous measurement. \\ 
\midrule
\multirow{-3.3}{*}{$\sqrt{F_{\omega}}\Delta \omega_{\text{est}} \geq 1$}   & 
This is the  Cramér-Rao bound for frequency estimation constrained by  $\Delta t. $  
Since $ \sqrt{F_{\omega}} \leq 2\Delta t $, it resembles TF "uncertainty relation" but is not truly one, as it represents only a constraint rather than a proper measure of uncertainty. The inequality is saturated by "dispersion-free" states, where $  \frac{\partial}{\partial \omega} \left[\frac{\Psi^*(\omega) }{ \Psi(\omega)}\right]=0 . $   \\ 
\end{tabular}
\end{ruledtabular}
\end{table*}

For the sake of completeness, let us also elaborate   \corr{on the estimation of the time variable. Notice first that thoughts inspired by  (\ref{Fisher_omega})  just interchanging the role of $t$ and $\omega $ cannot be repeated for the estimation of $\Delta t^2 $ since $\Delta \omega^2 $ is divergent. }  However this  issue can be corrected  by considering the variable $\sin t$ instead  and transformation induced by the generator $L$  in accordance with the commutator $[E,L]= E.$  The CR inequality obtains the simple form 
\begin{eqnarray}
   \Delta\sin^{2}t \ge 
   \frac{\langle \cos t \rangle ^2}{4  \Delta L^2},
\end{eqnarray}
which is nothing else than the uncertainty relation of quantum rotor (\ref{eq:tf_uncertainty})  derived here from informational principles.
We therefore conclude that the only framework permitting the simultaneous detection of TF variables must be based on the canonical variables of the quantum rotor.


\section{Simultaneous measurement of angular and angular momentum-like variable}\label{sec:simultaneous_measurement}

Clarifying the uncertainties, we are ready to formulate the problem of  simultaneous measurement in TF domain.
Extremal states of the uncertainty relations (\ref{eq:tf_uncertainty}) and (\ref{eq:normalized_tf_uncertainty}) are over-complete and span the closure relation in the form of a positive-operator-valued measure (POVM)

\begin{equation}
    \sum_{m\in\mathbb{Z}}\int_{-\pi}^{\pi}\frac{d\phi}{2\pi}|m,\phi\rangle \langle m,\phi|= \mathbb{1},
    \label{closure}
\end{equation}
where $|m,\phi\rangle$ are von Mises states, Eq.~(\ref{eq:von_mise_state}) with $n$ and $\alpha$ replaced with $m$ and $\phi$. The new parameters $m$ and $\phi$ have been introduced to distinguish the labels of the von Mises POVM elements, which coincide with outcomes of the considered simultaneous measurement, from the parameters $n$ and $\alpha$ labeling von Mises quantum states.

\begin{figure}
    \centering
    \includegraphics[width=\columnwidth] {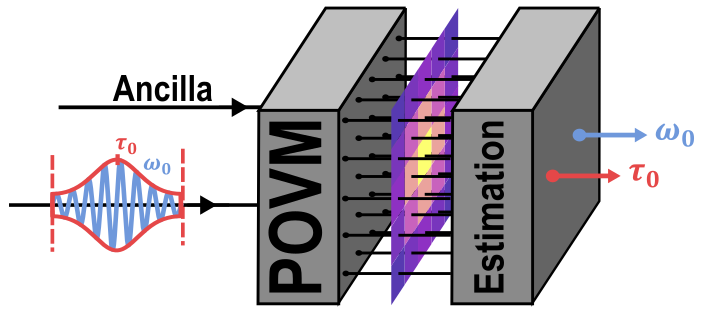}   
    \caption{Schematic of the realization of the required POVM. An ancillary field in the von Mises fiducial state $|0,0\rangle$ sets the POVM element which is applied to the signal. Measuring the output of each POVM element then allows for the simultaneous estimation of arrival time $\tau_0$ and carrier frequency $\omega_0$ of the signal with optimal precision.}
    \label{fig:POVM}
\end{figure}

\pad {Optimality stems from the representation of commuting operators  acting on composed systems of signal and ancillary spaces $\mathcal{H}_s\otimes\mathcal{H}_a$, namely
\begin{equation}
    \mathcal{L} = L_s + L_a, \: \mathcal{E} = E_s E_a^\dagger
    \label{eq:pair}
\end{equation}
with a corresponding commutation relation
\begin{equation}
    \left[\mathcal{L}, \mathcal{E}\right] = 0.
\end{equation}
Here\PL{, the} ancillary degrees of freedom represent \PL{the settings} of the apparatus \PL{which allows for measuring} the signal. When the ancillary degrees are traced, the measurement of the signal is represented by a POVM, as sketched symbolically  in Fig. \ref{fig:POVM}.  The detailed analysis  of this problem  for the rotor variables can be found in \cite{Mista_24}. 
}

Generalizing the sine operator (\ref{S}) to the signal-ancilla system, $\mathcal{S}=( e^{i\mathrm{arg}\langle\mathcal{E}\rangle}\mathcal{E}^\dagger-e^{-i\mathrm{arg}\langle\mathcal{E} \rangle}\mathcal{E})/2i$, eliminating an unwanted influence of the ancilla by imposing the constraints $\arg \langle E_a \rangle  = 0$ and $\arg \langle E_a^{2} \rangle  = 0$, 
and taking into account  independence of the signal and the ancilla, the variances for the simultaneous measurement of the commuting pair (\ref{eq:pair}) can be cast in the form \cite{Mista_22} 

\begin{align}
    \Delta\mathcal{L}^2&=\Delta L_s^2 + \Delta L_a^2,\\
    \Delta\mathcal{S}^2&=|\langle E_a^2 \rangle|\Delta S_s^2+\Delta S_a^2.
\end{align}

\PL{Here, the} simultaneous detection introduces added noise in both the angular-momentum-like and the angular variable. 
The latter, however, is suffering from an additional multiplicative factor $|\langle E_a^2 \rangle|$. 
This is a consequence of the fact that the angular variable is always measured with respect to some reference.
As a result, the structure of the optimal simultaneous measurement is more complex than in the case of uncertainty relations derived from a Heisenberg algebra, c.f. \cite{Mista_24}.

\begin{figure}
    \centering
    \includegraphics[width=\columnwidth]{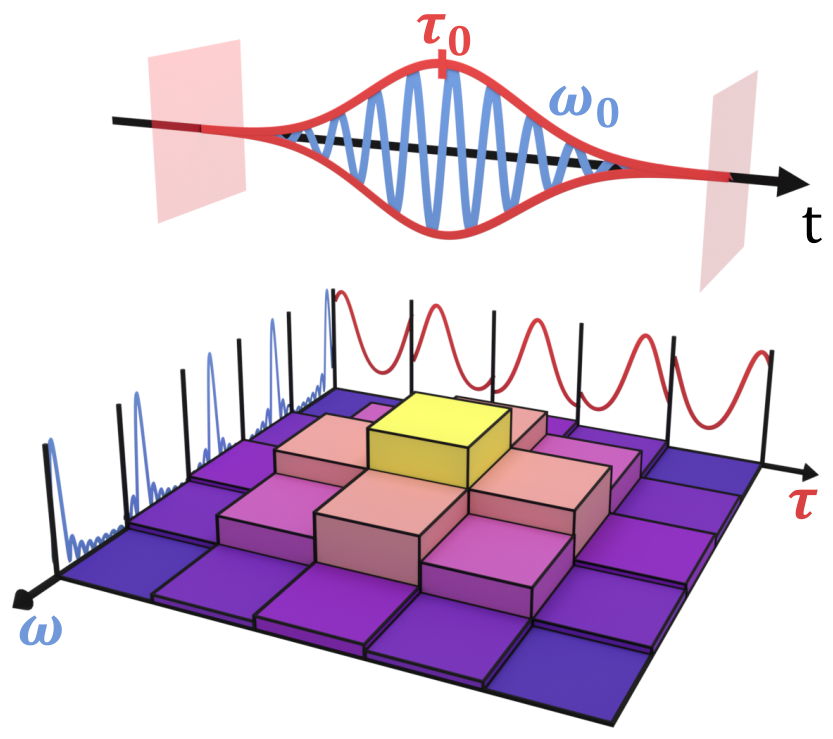}
    \caption{The signal is projected onto temporally and spectrally shifted von Mises states, sketched in the rows and columns of the matrix. For each combination of shifts, the complex field overlap between the signal and the projection mode is measured, which yields the signal $Q$-function in the basis of von Mises states.}
    \label{fig:Q-function}
\end{figure}

In the following, we will consider a situation in which a signal that is prepared in a von Mises state with a specific spread parameter $\kappa$ is projected onto a sequence of von Mises states with equal $\kappa$  but shifted in time and frequency variables, as outlined in Fig. \ref{fig:Q-function}. For this scenario, we find the following explicit expressions for the uncertainty relations
\begin{align}
    \Delta L\,\Delta S&= \frac12\frac{I_1(2\kappa)}{I_0(2\kappa)}, \label{eq:explicit_state_1}\\
    \Delta \mathcal{L}\,\Delta \mathcal{S}&=\frac{I_1(2\kappa)}{I_0(2\kappa)}\sqrt{\frac12\left[1+\frac{I_2(2\kappa)}{I_0(2\kappa)}\right]},\label{eq:explicit_measurement_1}
\end{align}
where (\ref{eq:explicit_state_1}) pertains to the intrinsic uncertainty of the signal state and (\ref{eq:explicit_measurement_1}) to the simultaneous measurement of angular and angular-momentum-like variable of the signal, respectively. \sugg{ The modified Bessel functions \(I_n(z)\) \cite{Watson_44} appearing here arise naturally as Fourier moments of the von Mises angular distribution \(\exp[2\kappa \cos(\phi-\alpha)]/2\pi I_{0}(2\kappa)\). Consequently, the ratios of modified Bessel functions entering the uncertainty relations directly reflect the relative weights of these moments. For a brief review of useful identities and properties of modified Bessel functions in this context, see Appendix~A in Ref.~\cite{Mista_22}.
 }

For the normalized uncertainties, we obtain
\begin{align}
    \Delta L\,\sigma&= \frac{1}{2}, \label{eq:explicit_state_2}\\
    \Delta\mathcal{L}\,\Sigma&=\frac{I_0(2\kappa)}{I_1(2\kappa)}\sqrt{\frac12\left[1+\frac{I_2(2\kappa)}{I_0(2\kappa)}\right]}\label{eq:explicit_measurement_2},
\end{align}
\PL{where we have defined $\Sigma\equiv\Delta\mathcal{S}/ (|\langle E_s \rangle|\,|\langle E_a \rangle|)$ and}, again, (\ref{eq:explicit_state_2}) refers  to the signal state and (\ref{eq:explicit_measurement_2}) to the simultaneous measurement, respectively.  Note that the given limit for simultaneous measurement corresponds to a strategy where the state with a given spread parameter $\kappa$ is projected onto a system of states with the same $\kappa$. This would be optimal for a system of harmonic oscillators, but it is slightly suboptimal for pulses. However, these deviations are not significant for the present purposes; further details and exact bound can be found in Ref. \cite{Mista_24}.

As a final step, we answer the question \PL{of} how to extract the uncertainties associated with the signal state (Eqs. (\ref{eq:explicit_state_1}) and (\ref{eq:explicit_state_2})) from the measurement that suffers from unavoidable additional noise. Because the detection based on the POVM (\ref{closure}) is over-complete, we may use the measurement results to estimate the signal state via quantum state tomography. In fact, projections of the signal state $\rho$ into von Mises states are sampling the $Q$-function \cite{Mista_22}, which is defined as
\begin{equation}
    Q(m,\phi) = \frac{1}{2\pi}\langle m,\phi|\rho|m,\phi\rangle.
\label{Q}
\end{equation}
This is sketched in Fig.~\ref{fig:Q-function}.
From the measured $Q$-function, the signal state $\rho$ can be reconstructed using methods such as MaxLik reconstruction \cite{Hradil_1997,Hradil_2004}. While the $Q$-function is influenced by both the signal state $\rho$ and the measurement, it can be transformed into the Wigner function, which isolates the uncertainties associated specifically with the quantum state. The phase space representation of rotor-like systems will be explored in more detail elsewhere, but for the purposes of this paper, we will use the formalism developed for vortex beams in \cite{Rigas_11}. From this transformation, the uncertainty of the underlying signal state $\rho$ can be extracted. However, it is important to note that these uncertainties are not directly measured but rather estimated.

\section{Experimental implementation}
\label{sec_Experiment}

The key to an experimental implementation of the optimal simultaneous estimation of the arrival time and carrier frequency of a time-dependent signal is the implementation of the POVM (\ref{closure}). 
We stated in Sec. \ref{sec:simultaneous_measurement} that we require an ancillary system that corresponds to a specific setting of the measurement instrument.
As a matter of fact, this translates to projecting the signal into a set of temporally and spectrally shifted von Mises states, characterized by $|m,\phi\rangle=D(m,\phi)|0,0\rangle$, c.f. Eqs. (\ref{eq:von_mise_state}) and (\ref{closure}) and Fig. \ref{fig:POVM}.

In the experiment, we used a so-called quantum pulse gate (QPG) \cite{eckstein_quantum_2011,brecht_demonstration_2014} as the measurement device.
The QPG is based on dispersion-engineered sum-frequency generation (SFG) in a periodically poled lithium niobate waveguide and implements the projection of an input signal onto a user-defined temporal mode that is, a complex time-frequency distribution \cite{ansari_temporal-mode_2017,de_realization_2024}.
To this end, we shape the complex spectrum of a pump pulse driving the QPG into a set of spectrally and temporally shifted von Mises states. 
\pad {Measuring the converted output intensity after the QPG quantifies the complex field overlap between the signal and the projection mode, directly corresponding to the output of the POVM element, in close analogy with the theoretical formulation.} 
In the experiment, the pump takes on the role of ancillary system that sets the state of the measurement device. 

\begin{figure}
    \centering
    \includegraphics[width=\columnwidth]{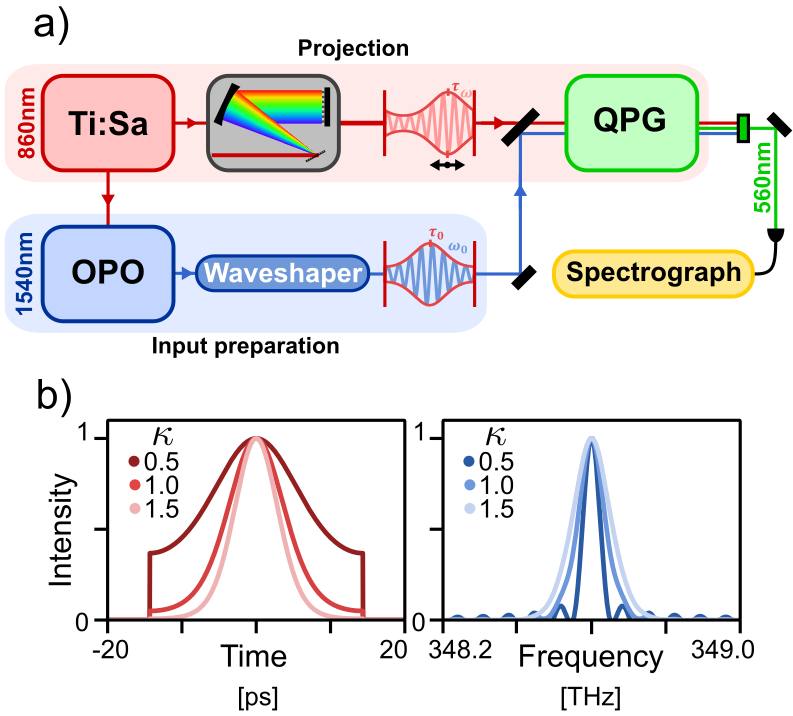}
        \caption{a) Schematic depiction of the experimental setup. The measured signal pulse is generated by shaping it's complex spectrum via a commercial waveshaper. This signal is  projected into varying von-Mises states (different central frequencies and times) via the QPG to implement the complete POVM. The converted SFG light from the QPG is detected and filtered via a spectrograph. \PL{b) Time and frequency depiction of von Mises states as they were used in the experiment.} 
    }
    \label{scheme}
\end{figure}

\begin{figure*}[ht]
\begin{center}
\includegraphics[]{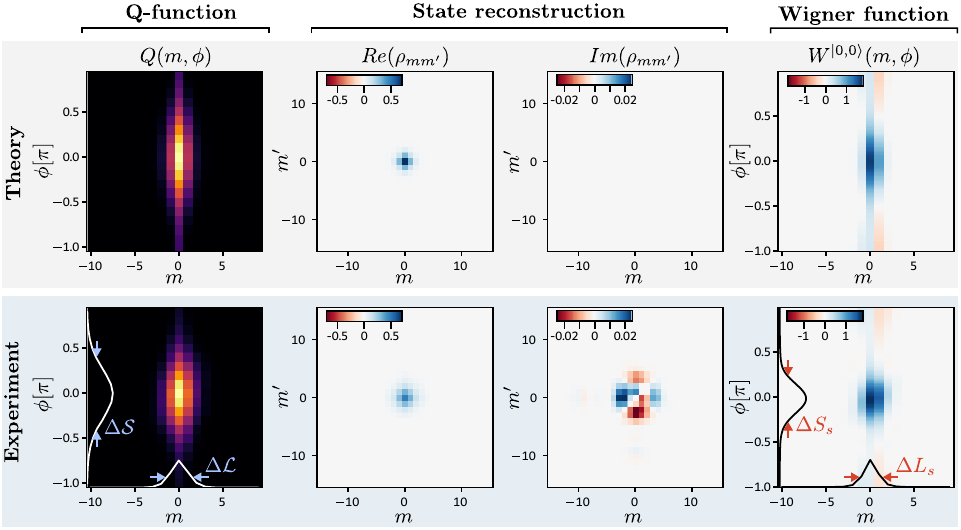}
\end{center}
\caption{Quantum state tomography of the fiducial signal von Mises state. \corr{  In the experiment   state with  $m=10000$ was used as fiducial state. Since just the relative difference in  $m$ matters as well as angular variable matters, we will  keep the notation  as   $|0,0\rangle$   with $\kappa=1$ centered at the origin of the time interval. } From the measured $Q$-function (lower left), the time-frequency sector of the components $\rho_{mm'}$ is reconstructed via an iterative MaxLik algorithm. It's real and imaginary part are plotted in the central columns. \PL{For the reconstructed state 
 $\rho_{mm'}$ we obtain the Wigner function (lower right).} The upper row shows the theory for the respective quantities. See text for details.}
\label{fig:Q-function-measurement}
\end{figure*}
Fig. \ref{scheme} shows a schematic of the experimental setup.
We generate the pump field for the QPG from an ultrafast Ti:Sapph oscillator that generates pulses at a central wavelength of $860\,$nm with a repetition rate of $80\,$MHz.
The spectrum of the pump pulses is shaped into a von Mises profile with an in-house built pulse shaper based on a spatial light modulator in a 4-f-line \cite{monmayrant_newcomers_2010} with a resolution of  $\delta \nu_{shaper,p} = 10\, \text{GHz}$. 
A part of the Ti:Sapph oscillator pumps an optical parametric oscillator (OPO), which generates signal pulses with a central wavelength of 1540 nm. 
Their spectrum is subsequently shaped into von Mises profile using a commercial wave shaper with a resolution of  $\delta \nu_{shaper,in} = 1\,\text{GHz}$.
 We created 12 signal states with spread values $ \kappa$ in the range $\kappa\in\left[0.2, 8\right].$
We chose the duration of the time interval as $T=28.6\,$ps, while we set \sugg{$\tau_0 =0$} and  $m=10000.$ 
The latter was chosen because the modes with orders in $m\in\left[9900, 10100\right]$ were compatible with the total shaping range of $14\,$nm of the pump pulse shaper.
We set the power of the pump beam to $0.5\,$mW and the power of the signal beam to $1.8\,\mu$W, where the latter corresponds to around 17700 photons per pulse.
Thereafter, we overlap signal and pump pulses on a dichroic mirror and couple them into the QPG sample.
The sample has a length of $40\,$mm and a poling period of $4.32 \mu\text{m}$ and is heated to a working temperature of  $160^{\circ}\text{C}$ to achieve optimal QPG operation and to prevent photorefraction. 
We separated the upconverted output of the QPG from the residual input fields with another dichroic mirror and detected it with a commercial single-photon sensitive CCD spectrograph. 
We set an integration time of $100\,$ms for each POVM element to ensure good statistical accuracy of the detected signal. For instance, typical data for $\kappa =1$ yield detected signal counts ranging from $886$ to $53324, $ with the total number of counts for the scanned $Q$-function of the state reaching approximately $1.47 \times 10^6.$
Although the detected data fluctuate, the maximum count values remain on the order of tens of thousands, while the total number of counts varies between $1.0 \times 10^6$ and $1.8  \times 10^6$. For theoretical analysis, the data are filtered by subtracting the background.
 In total, we projected each signal state onto 400 von Mises states by choosing 20 pump delays $\tau_p\in\left[-T/2, T/2\right]$ and 20 frequency shifts (characterized by  discrete $m$),  shaping the pump pulses according to each possible combination. 
The measurements were repeated for 12 different values of $\kappa$, which characterizes the  spread in time and frequency of the von Mises functions, ensuring that both signal and pump pulses had the same width.

\section{Results and Discussion}\label{sec_Results}

\pad {In the experiment, the $Q-$function of the signal, prepared in the von Mises state, is sampled according to Eq.~(\ref{Q}), mapping points in the TF phase space. The \PL{lower left} panel of Fig. \ref{fig:Q-function-measurement} presents the measured data for a signal state \PL{$\rho$} with a spread parameter $\kappa=1$, centered at the origin of the time interval ($\tau_0=0$), corresponding to the fiducial von Mises state $|0,0\rangle. $  From this measured $Q$-function, we extract the TF sector of the signal state, which is an attenuated coherent pulse, showing Poissonian photon number statistics. This retrieval is performed using an iterative maximum likelihood algorithm, and the resulting data is plotted in the \PL{central columns} of Fig. \ref{fig:Q-function-measurement}. To further analyze the intrinsic uncertainties  of TF variables of the signal state (cf. Eqs. (\ref{eq:explicit_state_1}) and (\ref{eq:explicit_state_2})), we compute \PL{the Wigner function \cite{Rigas_11}, shown in the right column.} The \PL{upper} row displays the theoretical values for these quantities.
}

 The uncertainties in the detection process are directly derived from the marginal distributions of the measured $Q$-functions, whereas the uncertainties of the signal state itself are obtained from the marginal distributions of the reconstructed Wigner functions.
\begin{figure}
    \centering
    \includegraphics[width=0.90\columnwidth]{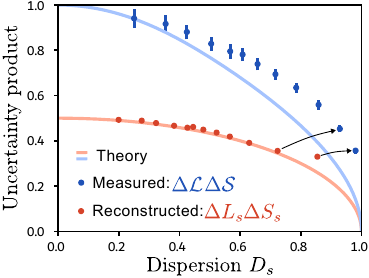} 
    \caption{Uncertainty products $\Delta L_{s}\,\Delta S_{s}$, Eq.~(\ref{eq:explicit_state_1}), \PL{(orange)} and $\Delta \mathcal{L}\,\Delta \mathcal{S}$, Eq.~(\ref{eq:explicit_measurement_1}), \PL{(blue)} obtained from projecting signals in von Mises states onto von Mises states with similar dispersion $D_s$. The experimentally retrieved states \PL{(orange circles)} saturate the quantum bound  \PL{(orange line)}, while the uncertainty in the joint estimation of arrival time and carrier frequency  \PL{(blue circles)} is slightly above the theoretical limit  \PL{(orange line)}, which is a consequence of experimental imperfections.}
    \label{fig:mu}
\end{figure}

\begin{figure}
    \centering
    \includegraphics[width=0.97\columnwidth]{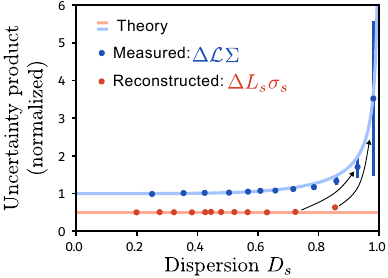} 
    \caption{Normalized uncertainty products $\Delta L_{s}\,\sigma_{s}$, Eq.~(\ref{eq:explicit_state_2}),  \PL{(orange)} and $\Delta\mathcal{L}\,\Sigma$, Eq.~(\ref{eq:explicit_measurement_2}),  \PL{(blue)}. The normalization mitigates the effects of experimental imperfections such that both the states and simultaneous measurements now saturate the respective bound, however at the cost of increased error bars in the case of the simultaneous measurement.}
    \label{fig:mu2}
\end{figure}

Fig.~\ref{fig:mu} \PL{shows} the non-normalized uncertainty products (\ref{eq:explicit_state_1}) and (\ref{eq:explicit_measurement_1}). The ultimate bound pertaining to the signal state is plotted as a blue line for different values of the signal spread parameter $\kappa_{s}$. For convenience, we plot on the horizontal axis the statistical dispersion \cite{Holevo_11} of the signal state, 
$D_{s}=\sqrt{1-I^2_1(2\kappa_{s})/I^2_0(2\kappa_{s})}$ \cite{Mista_24}, which is a monotonically decreasing function of $\kappa_{s}$ mapping the semi-infinite interval $[0,+\infty]$ onto the finite interval $[0,1]$. Thus, for instance, $D_{s}=1$ corresponds to $\kappa_{s}=0$ when the signal state reduces to the angular momentum eigenstate $|n=0\rangle_{s}$, whereas $D_{s}=0$ is reached in the limit $\kappa_{s}\rightarrow+\infty$ when the signal state approaches the eigenstate of the shift operator $|\phi=0\rangle_{s}$.

The corresponding data retrieved from the reconstructed Wigner function are plotted as orange circles. The error bars originate from Monte Carlo simulations which use the statistical distribution of the available data (parametric bootstrap).  We can see that the signal states saturate the bound for most of the values of the spread parameter. In contrast to this, the minimum uncertainty bound for the simultaneous measurement is plotted as a \PL{solid blue line}, with the corresponding data being \PL{blue circles}.  The black arrows indicate corresponding points. Due to the noise addition during measurement, individual points move to the right and upward in the plot.  The primary reason for this misalignment is the complexity of the detection process itself, which requires projections onto well-defined states that can never be implemented with perfect precision. As a result, systematic errors appear as offsets between the presumed and estimated states.  In our interpretation, this offset represents an inevitable  error in the time  variable. Together with the error bars in $Q$ detection, this allows us to quantify the uncertainties of the complementary pair. This provides a practical way to address an otherwise intractable and often overlooked issue in quantum tomography: how to assign error bars to an estimated quantum state. First, we find that the minimum uncertainty for the measurement is larger than that of the state due to the added noise. Second, we also find that the data poorly  saturate the bound for  squeezing values $\kappa$ close to zero. 
\sugg{  As can be seen from the description of the experiment, both the signal states and the measurement projectors are emulated using bandwidth-limited optical pulses, which necessarily leads to approximate realizations of the ideal von Mises profiles. As a result, states with sharp temporal features may exhibit small deviations, leading to minor discrepancies between theoretical predictions and experimentally retrieved uncertainties.}

\sugg{ These deviations are not fundamental. In physical systems with native rotor symmetry, such as resonator-based or periodically constrained systems, the frequency spectrum is inherently discrete and von Mises states arise naturally as minimum-uncertainty states without requiring unbounded bandwidth. Our experiment is therefore intended as a proof-of-principle demonstration of the theoretical framework rather than as a resource-optimized implementation.
}

Another possible source of deviations is a systematic variation in the signal intensity, which may occur for different values of the spread parameter $\kappa$. This seems a reasonable assumption when one recognizes that spectral resolution is limited but large spread corresponds to sharp spectral features of the von Mises distribution.
 To test this, Fig. \ref{fig:mu2} plots the normalized uncertainties (\ref{eq:explicit_state_2}) and (\ref{eq:explicit_measurement_2}), which we expect to be less sensitive to variations in the signal intensity. Indeed, we find that in this representation the data lie closer to both the minimum uncertainty bounds for the quantum state and the simultaneous measurement.

\sugg{Finally, we clarify the role of von Mises states in our framework. The present experiment addresses their TF uncertainties but does not consider wave propagation through dispersive media. The parameter 
$\kappa$  quantifies the relative spread of complementary variables and plays a role analogous to a squeezing parameter for Gaussian states. In general, 
it is not invariant under dispersive propagation, as dispersion broadens an initially optimal pulse, increasing the uncertainty product and degrading optimality. While the qualitative behavior is clear, a quantitative description of this evolution lies beyond the scope of the present manuscript.}

\section{Conclusion}\label{sec_Conclusion}
 { We have formulated a theoretical framework for time and frequency as conjugate variables in quantum mechanics. The standard approach, based on analogy with Heisenberg algebra, fails when time is constrained to a finite domain, as the variance of the time variable is not a suitable measure and the variance of frequency diverges. These issues can be resolved using a quantum rotor model characterized by the Euclidean $E(2)$ algebra.
Our novel theoretical concept was applied to optical pulses  introducing TF uncertainty relations that are saturated by over-complete von Mises states, which play a role analogous to that of squeezed states in quantum optics. Moreover, the developed theory enables us to formulate the problem of simultaneous detection of TF variables, allowing us to fully exploit their complementary nature. }

 { We experimentally validated our theoretical predictions by shaping a test signal into various von Mises states with different spread parameters in the time domain. These states were projected onto a series of von Mises states using our quantum pulse gate, which is designed to project onto arbitrary, user-defined time-frequency superposition states. Our measurements directly sampled the $Q-$function of the signal states, from which we reconstructed the underlying states by  maximum likelihood estimation. A detailed analysis of our results confirms two key findings: (i) we can prepare signals in the TF domain that closely adhere to the uncertainty bound, and (ii) we can perform quantum-limited measurements. We successfully reconstructed the Wigner function of the underlying state with quantum-limited precision, a feat not previously demonstrated in quantum rotor-like systems. 
 }

\sugg{While our work is primarily focused on establishing a fundamental theoretical framework, it naturally connects to a class of advanced optical and quantum metrology problems. In particular, scenarios in which both arrival time and frequency shift must be inferred simultaneously within a finite detection window and under limited resources provide a natural context for our results.}

\sugg{ Examples include tracking and ranging tasks in which time-of-flight and Doppler information are both relevant and cannot be optimally treated as independent variables. In such situations, a simultaneous measurement strategy that respects the correct TF uncertainty structure may offer a principled way to distribute the available quantum information between the two parameters.} 

\sugg{ More generally, our results suggest that optimal TF detection schemes need not rely on a fixed choice of state basis. Instead, both the signal preparation and the subsequent measurement projections can be tailored to the specific boundary conditions and information-theoretic objectives of a given task. This perspective emphasizes the role of adaptive, application-driven strategies for exploiting complementary variables, beyond the von Mises basis considered here.
The physical implications of this approach extend beyond the optical domain to other systems described by the 
$ E(2) $
 algebra. A notable example is provided by circuits with Josephson junctions, where the complementary variables correspond to the number of Cooper pairs tunneling through the barrier and an exponential operator of the superconducting phase. In such systems, the choice of optimal states and measurements is likewise dictated by the underlying algebra and boundary conditions, highlighting the generality of the framework developed in this work. }

\section*{Acknowledgment}
This project has received funding from the European Union’s Horizon Europe research and
innovation programme under grant agreement No 899587 (STORMYTUNE).

\section*{Disclosures}
The authors declare no conflicts of interest.

\bibliography{bibOlomouc}

@article{Altaie_22,
author = {Altaie, Mohammed Basil and Hodgson, Daniel and Beige, Almut},
year = {2022},
month = {06},
pages = {897305},
title = {Time and Quantum Clocks: A Review of Recent Developments},
volume = {10},
journal = {Front. Phys.},
doi = {10.3389/fphy.2022.897305}
}

@article{Bonneau_01,
  title={Self-adjoint extensions of operators and the teaching of quantum mechanics},
  author={Bonneau, Guy and Faraut, Jacques and Valent, Galliano},
  journal={Am. J. Phys.},
  volume={69},
  number={3},
  pages={322--331},
  year={2001},
  publisher={American Association of Physics Teachers}
}

@article{Bhattacharjee2025,
author = {Abhinandan Bhattacharjee and Patrick Folge and Laura Serino and Jaroslav \v{R}eh\'{a}\v{c}ek and Zden\v{e}k Hradil and Christine Silberhorn and Benjamin Brecht},
journal = {Opt. Express},
number = {3},
pages = {5551--5561},
publisher = {Optica Publishing Group},
title = {Pulse characterization at the single-photon level through chronocyclic {Q}-function measurements},
volume = {33},
month = {Feb},
year = {2025},
url = {https://opg.optica.org/oe/abstract.cfm?URI=oe-33-3-5551},
doi = {10.1364/OE.540125},
}

@article{Brecht2015,
  title = {Photon Temporal Modes: A Complete Framework for Quantum Information Science},
  author = {Brecht, B. and Reddy, Dileep V. and Silberhorn, C. and Raymer, M. G.},
  journal = {Phys. Rev. X},
  volume = {5},
  issue = {4},
  pages = {041017},
  numpages = {17},
  year = {2015},
  month = {Oct},
  publisher = {American Physical Society},
  doi = {10.1103/PhysRevX.5.041017},
  url = {https://link.aps.org/doi/10.1103/PhysRevX.5.041017}
}

@article{Mista_24,
  title={Unifying uncertainties for rotorlike quantum systems},
  author={Mi{\v{s}}ta, Jr., Ladislav and Mi{\v{s}}ta, Matou{\v{s}} and Hradil, Zden{\v{e}}k},
  journal={Phys. Rev. A},
  volume={110},
  number={3},
  pages={032208},
  year={2024},
  publisher={APS}
}

@book{Helstrom_76,
  title={Quantum detection and estimation theory.},
  author={Helstrom, C. W.},
  chapter={VIII},
  pages={277},
  year={1976},
  publisher={Academic Press},
  address = {New York}
}

@book{Holevo_11,
  title={Probabilistic and Statistical Aspects of Quantum Theory},
  author={Holevo, Alexander Semenovich},
  series={Publications of the Scuola Normale Superiore},
  year={2011},
  publisher={Edizioni della Normale},
  address={Pisa},
  edition={2nd}
}

@article{Einstein_35,
  title={Can quantum-mechanical description of physical reality be considered complete?},
  author={Einstein, Albert and Podolsky, Boris and Rosen, Nathan},
  journal={Phys. Rev.},
  volume={\textbf{47}},
  number={10},
  pages={777},
  year={1935},
  publisher={APS}
}

@article{Glauber_63,
  title={Coherent and incoherent states of the radiation field},
  author={Glauber, Roy J},
  journal={Phys. Rev.},
  volume={\textbf{131}},
  pages={2766--2788},
  year={1963}
}

@article{Arthurs_65,
  title={{B.S.T.J.} briefs: On the simultaneous measurement of a pair of conjugate observables},
  author={Arthurs, E and Kelly, J. L.},
  journal={Bell Syst. Tech. J.},
  volume={\textbf{44}},
  number={4},
  pages={725--729},
  year={1965},
  publisher={Nokia Bell Labs}
}

@book{Watson_44,
  edition={2nd ed.},
  title={A Treatise on the Theory of Bessel Functions},
  author={Watson, GN},
  publisher={Cambridge University Press, Cambridge, UK},
  year={1944}
}

@article{Rigas_11,
  title={Orbital angular momentum in phase space},
  author={Rigas, I. and Sánchez-Soto, L. L. and Klimov, A. B. and Řeháček, J. and Hradil, Z.},
  journal={Ann. Phys.},
  volume={\textbf{326}},
  pages={426--439},
  year={2011},
  publisher={Elsevier}
}

@article{Plebanski_00,
title = "Remarks on deformation quantization on the cylinder",
author = "Pleba{\'n}ski, {J. F.} and M. Przanowski and J. Tosiek and Turrubiates, {F. J.}",
year = "2000",
volume = "\textbf{31}",
pages = "561--587",
journal = "Acta Phys. Pol. B",
}

@article{Kastrup_06,
  title={Quantization of the canonically conjugate pair angle and orbital angular momentum},
  author={Kastrup, H. A.},
  journal={Phys. Rev. A},
  volume={\textbf{73}},
  pages={052104},
  year={2006},
  publisher={APS}
}

@article{Husimi_40,
  title={Some formal properties of the density matrix},
  author={Husimi, K.},
  journal={Proc. Phys. Math. Soc. Jpn.},
  volume={\textbf{22}},
  pages={264--314},
  year={1940},
  publisher={THE PHYSICAL SOCIETY OF JAPAN, The Mathematical Society of Japan}
}

@article{Wigner_32,
  title={On the quantum correction for thermodynamic equilibrium.},
  author={Wigner, E. P.},
  journal={Phys. Rev.},
  volume={\textbf{40}},
  pages={749-759},
  year={1932},
}

@article{Bluhm_95,
  title={Elliptical squeezed states and {R}ydberg wave packets},
  author={Bluhm, Robert and Kosteleck{\'y}, V Alan and Tudose, Bogdan},
  journal={Phys. Rev. A},
  volume={\textbf{52}},
  number={3},
  pages={2234},
  year={1995},
  publisher={APS}
}

@article{Mista_22,
  title={Angle and angular momentum: Uncertainty relations, simultaneous measurement, and phase-space representation},
  author={Mi{\v{s}}ta, Jr., L. and de Guise, H. and {\v{R}}eh{\'a}{\v{c}}ek, J. and Hradil, Z.},
  journal={Phys. Rev. A},
  volume={106},
  pages={022204},
  year={2022},
  publisher={APS}
}

@article{Hradil_10,
  title={Angular performance measure for tighter uncertainty relations},
  author={Hradil, Z. and {\v{R}}eh{\'a}{\v{c}}ek, J. and Klimov, A. B. and Rigas, I. and S{\'a}nchez-Soto, L. L.},
  journal={Phys. Rev. A},
  volume={81},
  pages={014103},
  year={2010},
  publisher={APS}
}

@article{Albert_17,
  title={General phase spaces: from discrete variables to rotor and continuum limits},
  author={Albert, Victor V and Pascazio, Saverio and Devoret, Michel H},
  journal={J. Phys. A: Math. Theor.},
  volume={50},
  number={50},
  pages={504002},
  year={2017},
  publisher={IOP Publishing}
}

@article{Heisenberg_27,
  title={{\"U}ber den anschaulichen Inhalt der quantentheoretischen Kinematik und Mechanik},
  author={Heisenberg, Werner},
  journal={Z. Phys.},
  volume={43},
  number={3},
  pages={172--198},
  year={1927},
  publisher={Springer}
}

@article{Robertson_29,
  title={The uncertainty principle},
  author={Robertson, Howard Percy},
  journal={Phys. Rev.},
  volume={34},
  number={1},
  pages={163},
  year={1929},
  publisher={APS}
}

@article{Yuen_76,
  title = {Two-photon coherent states of the radiation field},
  author = {Yuen, Horace P.},
  journal = {Phys. Rev. A},
  volume = {13},
  issue = {6},
  pages = {2226--2243},
  numpages = {0},
  year = {1976},
  month = {Jun},
  publisher = {American Physical Society},
  doi = {10.1103/PhysRevA.13.2226},
  url = {https://link.aps.org/doi/10.1103/PhysRevA.13.2226}
}

@article{Hradil_1997,
  title = {Quantum-state estimation},
  author = {Hradil, Z.},
  journal = {Phys. Rev. A},
  volume = {55},
  issue = {3},
  pages = {R1561--R1564},
  numpages = {0},
  year = {1997},
  month = {Mar},
  publisher = {American Physical Society},
  doi = {10.1103/PhysRevA.55.R1561},
  url = {https://link.aps.org/doi/10.1103/PhysRevA.55.R1561}
}

@incollection{Hradil_2004,
    author={Hradil, Zden{\v{e}}k and {\v{R}}eh{\'a}{\v{c}}ek, Jaroslav and Fiur{\'a}{\v{s}}ek, Jarom{\'\i}r and Je{\v{z}}ek, Miroslav},
    title={3 Maximum-likelihood methods in quantum mechanics},
    booktitle = {Quantum state estimation},
    chapter = {7},
    pages = {59--112},
    editor = {{\v{R}}eh{\'a}{\v{c}}ek, Jaroslav and Paris, Matteo},
    publisher = {Springer},
    year = {2004}
}

@article{Rehacek_04,
  title={Uncertainty relations from {Fisher} information},
  author={J. \v{R}eh{\'a}{\v{c}}ek and Z. Hradil},
  journal={J. Mod. Opt.},
  volume={51},
  number={6-7},
  pages={979--982},
  year={2004},
  publisher={Taylor \& Francis}
}

@article{monmayrant_newcomers_2010,
	title = {A newcomer's guide to ultrashort pulse shaping and characterization},
	volume = {43},
	issn = {0953-4075, 1361-6455},
	url = {https://iopscience.iop.org/article/10.1088/0953-4075/43/10/103001},
	doi = {10.1088/0953-4075/43/10/103001},
	number = {10},
	urldate = {2021-01-14},
	journal = {J. Phys. B: At. Mol. Opt. Phys.},
	author = {Monmayrant, Antoine and Weber, Sébastien and Chatel, Béatrice},
	month = may,
	year = {2010},
	pages = {103001},
}

@article{eckstein_quantum_2011,
	title = {A quantum pulse gate based on spectrally engineered sum frequency generation},
	volume = {19},
	issn = {1094-4087},
	url = {https://www.osapublishing.org/oe/abstract.cfm?uri=oe-19-15-13770},
	doi = {10.1364/OE.19.013770},
	number = {15},
	urldate = {2021-01-14},
	journal = {Optics Express},
	author = {Eckstein, Andreas and Brecht, Benjamin and Silberhorn, Christine},
	month = jul,
	year = {2011},
	pages = {13770},
}

@article{brecht_demonstration_2014,
	title = {Demonstration of coherent time-frequency {Schmidt} mode selection using dispersion-engineered frequency conversion},
	journal = {Phys. Rev. A},
    volume = {90},
	author = {Brecht, Benjamin and Eckstein, Andreas and Ricken, Raimund and Quiring, Viktor and Suche, Hubertus and Sansoni, Linda and Silberhorn, Christine},
	year = {2014},
	pages = {030302(R)},
}

@article{ansari_temporal-mode_2017,
	title = {Temporal-mode measurement tomography of a quantum pulse gate},
	volume = {96},
	issn = {2469-9926, 2469-9934},
	url = {https://link.aps.org/doi/10.1103/PhysRevA.96.063817},
	doi = {10.1103/PhysRevA.96.063817},
	number = {6},
	urldate = {2021-01-14},
	journal = {Phys. Rev. A},
	author = {Ansari, Vahid and Harder, Georg and Allgaier, Markus and Brecht, Benjamin and Silberhorn, Christine},
	month = dec,
	year = {2017},
	pages = {063817},
}

@article{de_realization_2024,
	title = {Realization of high-fidelity unitary operations on up to 64 frequency bins},
	volume = {6},
	issn = {2643-1564},
	url = {https://link.aps.org/doi/10.1103/PhysRevResearch.6.L022040},
	doi = {10.1103/PhysRevResearch.6.L022040},
	number = {2},
	urldate = {2024-05-23},
	journal = {Phys. Rev. Res.},
	author = {De, Syamsundar and Ansari, Vahid and Sperling, Jan and Barkhofen, Sonja and Brecht, Benjamin and Silberhorn, Christine},
	month = may,
	year = {2024},
	pages = {L022040},
}

@article{Mandelstam1945,
  author = {Mandelstam, L. and Tamm, I.},
  title = {The Uncertainty Relation Between Energy and Time in Non-relativistic Quantum Mechanics},
  journal = {J. Phys. (USSR)},
  volume = {9},
  number = {4},
  pages = {249--254}, 
  year = {1945},
  publisher = {Russian Academy of Sciences},
  url = {https://daarb.narod.ru/mandtamm/mt-eng.pdf}, 
}

@article{Parhizkar2015,
author = {Reza Parhizkar and Yann Barbotin and Martin Vetterli},
title = {Sequences with minimal time–frequency uncertainty},
journal = {Appl. Comput. Harmon. Anal.},
volume = {38},
number = {3},
pages = {452-468},
year = {2015},
issn = {1063-5203},
doi = {https://doi.org/10.1016/j.acha.2014.07.001},
url = {https://www.sciencedirect.com/science/article/pii/S1063520314000906},
}

\section*{Appendix:  Fourier transform of von Mises distribution }\label{Appendix_A}

Here we show that von Mises states are to a quantum rotor what Gaussian states are to a harmonic oscillator. We start by noting that the overlap of two von Mises states $|n,\alpha\rangle$ and $|n',\alpha'\rangle$ with the same parameter $\kappa$, Eq.~(\ref{eq:von_mise_state}), is given by \cite{Kastrup_06,Mista_22}
\begin{eqnarray}
| \langle n, \alpha| n', \alpha' \rangle |^2=2\pi Q^{|n',\alpha'\rangle}(n,\alpha) = \frac{ I^2_{n-n'} [2 \kappa \cos(\frac{\alpha-\alpha'}{2} )  ]}{ I_0^{2}( 2\kappa)}.\nonumber\\
\end{eqnarray}

Consider now the Fourier transformation of an operator $A(n,\alpha)$ of an integer $n$ and an angle $\alpha$ \cite{Plebanski_00}
\begin{eqnarray}
\label{FTalpha}
(\mathcal{F}A)(l,\phi)=\sum_{n\in\mathbb{Z}}\int_{-\pi}^{\pi}\frac{d\alpha}{2\pi} e^{i(l\alpha-\phi n)}A(n,\alpha).
\end{eqnarray}
One can then show  \cite{Mista_22} that the Fourier transformation of the projector onto von Mises state $|n,\alpha\rangle$ reads as
\begin{eqnarray}\label{vMFT}
\left(\mathcal{F}|n,\alpha\rangle\langle n,\alpha|\right)(l,\phi)&=&e^{il\frac{\phi}{2}}\frac{I_l\left[2\kappa \cos\left(\frac{\phi}{2}\right)\right]}{I_0(2\kappa)}D(l,\phi),
\nonumber\\
\end{eqnarray}
where $D(l,\phi)$ is the displacement operator. Hence one finds straightforwardly
\begin{eqnarray}\label{vMoverlapFT}
\left(\mathcal{F}| \langle n, \alpha| n', \alpha' \rangle |^2\right)(l,\phi)&=&e^{il\frac{\phi}{2}}\frac{I_l\left[2\kappa \cos\left(\frac{\phi}{2}\right)\right]}{I_0(2\kappa)}\nonumber\\
&&\times\langle n', \alpha'|D(l,\phi)| n', \alpha' \rangle\nonumber\\
&=&e^{i(l\alpha'-\phi n')}\frac{I_l^{2}\left[2\kappa \cos\left(\frac{\phi}{2}\right)\right]}{I_0^{2}(2\kappa)},
\nonumber\\
\end{eqnarray}
where in the second equality we used the formula \cite{Mista_22}
\begin{eqnarray}\label{vMDelement}
\langle n', \alpha'|D(l,\phi)| n', \alpha'\rangle&=&e^{i(l\alpha'-\phi n')}e^{-il\frac{\phi}{2}}\frac{I_{l}\left[2\kappa\cos\left(\frac{\phi}{2}\right)\right]}{I_0(2\kappa)}.\nonumber\\
\end{eqnarray}
Equation (\ref{vMoverlapFT}) then can be rewritten as
\begin{eqnarray}\label{QC}
\left[\mathcal{F}Q^{|n',\alpha'\rangle}(n,\alpha)\right](l,\phi)&=&C_{Q}^{|n',\alpha'\rangle}(l,\phi),
\nonumber\\
\end{eqnarray}
where
\begin{eqnarray}\label{QC1}
Q^{|n',\alpha'\rangle}(n,\alpha)&=&\frac{I^2_{n-n'} [2 \kappa \cos(\frac{\alpha-\alpha'}{2})]}{2\pi I_0^{2}( 2\kappa)}\nonumber\\
C_{Q}^{|n',\alpha'\rangle}(l,\phi)&=&e^{i(l\alpha'-\phi n')}\frac{I_l^{2}\left[2\kappa \cos\left(\frac{\phi}{2}\right)\right]}{2\pi I_0^{2}(2\kappa)}.
\nonumber\\
\end{eqnarray}
However, this is nothing but the relation between the $Q$-function $Q^{|n',\alpha'\rangle}(n,\alpha)$ and the corresponding characteristic function $C_{Q}^{|n',\alpha'\rangle}(l,\phi)$ of the von Mises state $|n',\alpha'\rangle$, which are both given by a square of the modified Bessel function. The von Mises states thus exhibit a property analogous to the Gaussian states — namely, its $Q$-function preserves its shape under the respective Fourier transformation.


\end{document}